 \documentclass[aps,preprint,amsmath,amssymb]{revtex4}
\usepackage{amssymb}
\usepackage{graphicx}
\usepackage{subfigure}
\usepackage{dcolumn}
\usepackage{bm}
\usepackage{hyperref}

\begin{document}
\title{Zero Energy anomaly in one-dimensional Anderson lattice with
exponentially correlated weak diagonal disorder}

\author{Zongguo Wang}
\affiliation{
State Key Laboratory of Theoretical Physics,
Institute of Theoretical Physics,
Chinese Academy of Science, Beijing 100190, People's Republic of China
} \email{wangzg@itp.ac.cn}
\author{Kai Kang}
\affiliation{School of Physics, Peking University, Beijing 100871,
People's Republic of China}
\author{Shaojing Qin}
\affiliation{
State Key Laboratory of Theoretical Physics,
Institute of Theoretical Physics,
Chinese Academy of Science, Beijing 100190, People's Republic of China
}
\author{Chuilin Wang}
\affiliation{China Center of Advanced Science and Technology,
P. 0. Box 8730, Beijing 100190, People's Republic of China}

\date{\today}
\begin{abstract}
We calculated numerically the localization length of one-dimensional
Anderson model with correlated diagonal disorder. For zero energy
point in the weak disorder limit, we showed that the localization length
changes continuously as the correlation of the disorder increases.
We found that higher order terms of the correlation must be included
into the current perturbation result in order to give the correct
localization length, and to connect smoothly the anomaly at zero
correlation with the perturbation result for large correlation.
\end{abstract}

\maketitle

\section{Introduction}
\label{sec:intro}

Electronic transport properties, the motion of electrons, in a
random potential are closely related to the phenomenon of
Anderson localization\cite{PAnderson1958}.
The phenomena of Anderson localization have been studied
in various fields including photonics\cite{caoh2010},
cold atoms \cite{AdLagendijk2009}, circuits \cite{lazo2011},
and DNA molecules \cite{zhangw2004,zhangw2010}.
Many accurate numerical approaches have been developed, by
the quantum transfer matrix renormalization group method
for finite temperature systems \cite{lpyang2009},
the density matrix renormalization group method
for interacting systems \cite{pshmitteckert1998}, and the integral
equation method for systems in the thermodynamic limit
\cite{kkang2010,kkang2011}, respectively.  In this work we will
study the zero energy behavior for the one-dimensional model
with correlated weak diagonal disorder.  We first extend the
numerical method we developed earlier in Ref.~\cite{kkang2010}
for uncorrelated disorder to correlated system. Our numerical method
was an application of the transfer matrix method \cite{jbpendry1994}
in localized phase in the thermodynamic limit.

In one-dimensional Anderson model\cite{PAnderson1958}
with diagonal disorder is described by,
\begin{equation}
\psi_{i-1}+\psi_{i+1}=(E-\epsilon_i)\psi_i, \label{eq:se}
\end{equation}
where hopping term is set to unity and
$\psi_i$ is the electron wavefunction at site-$i$.
${\epsilon_i}$ is the on-site energy with a certain type of
random distribution which satisfying an exponential correlation:
$\langle\epsilon_i^2\rangle=\sigma^2$ and
$\langle\epsilon_i \epsilon_j\rangle=\sigma^2\exp[-|i-j|/l_{cor}]$
for different sites.
$\sigma^2$ and $l_{cor}$ are the strength and correlation length
for the disordered on-site energy, respectively.  Uncorrelated
disorder is given by $l_{cor}\to 0$.
Recently the anomaly around the band edge $E=\pm 2$ has been carefully
investigated. \cite{gurevich2011} In the following we focus on
the zero energy anomaly with exponentially correlated diagonal disorder.

All the eigenstates are exponentially localized for one-dimensional
uncorrelated disordered systems.  \cite{EAbrahams1979} The  Lyapunov
exponent $\gamma$ is the inverse of the localization length.
It is well known that for the zero energy anomaly of the uncorrelated disorder system,
the Lyapunov exponent $\gamma$ is singular at $E=\sigma=0$.
\cite{HSchomerus2002,AStone1983}.
The physical picture behind was also clear\cite{BAltshuler2003}.
For a box distribution of uncorrelated disorder with width $W$
and height $1/W$, the perturbation result revealed that the Lyapunov
exponent depends only on energy $E$ and disorder strength $W$
\cite{DThouless1979}
\begin{equation}
\gamma=\frac{W^2}{96(1-E^2/4)}.\label{eq:sp}
\end{equation}
At the band center, another perturbation yielded
\cite{MKappus1981,BDerrida1984,FIzrailev1998}
\begin{equation}
\gamma=\frac{W^2}{105.045\cdots}.\label{eq:c}
\end{equation}
The standard variance of the disorder is $\sigma^2=W^2/12$.
In uncorrelated systems, order by order
perturbation expansion in $\sigma^2$ and $E/\sigma^2$
has been demonstrated \cite{kkang2011}.

For exponentially correlated disorder, the formula for the Lyapunov
exponent at finite energy and in the weak disorder strength limit
is given by\cite{FIzrailev2001}
\begin{equation}
\gamma = {\frac {\sigma^2 } {8 ({1-{\frac{E^2}{4}}})}} \cdot
         {\frac { \sinh{\frac{1}{  l_{cor}}}}
          {1+\cosh{\frac{1}{  l_{cor}}}-{\frac{E^2}{2}}}}
\label{eq:gmcor}.
\end{equation}
It is straight forward to take the uncorrelated limit $l_{cor}\to 0$ of
formula Eq.~(\ref{eq:gmcor}), then obtain $\gamma/\sigma^2=1/8$ when $E$
approaches to $0$, i.e.
${\displaystyle
\lim_{E\to 0}\lim_{\sigma^2\to 0} {\frac \gamma {\sigma^2}}=1/8 }$.
On the other hand, if we stay at $E=0$, we should have
$\gamma/\sigma^2=1/8.754$ in the uncorrelated limit in accordance to  Eq.~(\ref{eq:c}),
which implies
${\displaystyle
\lim_{\sigma^2\to 0} \lim_{E\to 0} {\frac \gamma {\sigma^2}}=1/8.754 }$.
Therefore, we found that the order of the limiting
processes for $E\to 0$ and $\sigma^2 \to 0$ can not be interchanged.
It means that the point $E=\sigma=0$ remains singular for perturbation
expansions in $\sigma^2$ and $E$ for correlated disorders.
The existence of strong anomalies phenomena in a correlated disorder
system was pointed by Titov and Schomerus\cite{HSchomerus2005}.
In this work we study the anomaly at $E=0$.

\section{Parametrization method}

In the transfer matrix method,  Eq.~(\ref{eq:se}) can be written as
\begin{equation}
\Psi_{i+1}=\left(\begin{array}{c} \psi_{i+1} \\ \psi_i
\end{array}\right)= \left(\begin{array}{cc} {E-\epsilon_i} & -1 \\ 1 & 0
\end{array}\right) \left(\begin{array}{c} \psi_i \\ \psi_{i-1}
\end{array}\right)=\mathbf{T}_i\Psi_i, \label{eq:tm}
\end{equation}
where $\mathbf{T}_i$ is the transfer matrix.

Using a parametrization method of the transfer matrix proposed in our
previous work \cite{kkang2010,kkang2011}, we will calculate the Lyapunov
exponent in the thermodynamic limit within the localization regime.
Let $\mathbf{M}_L=\mathbf{T}_L\mathbf{T}_{L-1}\cdots\mathbf{T}_1$.
Then we parameterize $\mathbf{M}\mathbf{M}^t$ as follows
\begin{equation}
\mathbf{U}(\theta_L)\mathbf{M}_L\mathbf{M}^t_L \mathbf{U}(-\theta_L)
=\left(\begin{array}{cc} e^{2\lambda_L} & \\ & e^{-2\lambda_L}
\end{array} \right),
\end{equation}
where $\mathbf{M}^t$ is the transpose of $\mathbf{M}$ and
\begin{equation}
\mathbf{U}(\theta_L)=\left(\begin{array}{cc} \cos\theta_L & -\sin\theta_L
\\ \sin\theta_L & \cos\theta_L \end{array} \right).
\end{equation}
The recursion relation of $\theta$ in the large $L$ limit is
\begin{equation}
\tan\theta_{L+1}=\frac{1}{E-\epsilon_{L+1}-\tan\theta_L}. \label{eq:theta}
\end{equation}

We introduce the correlations between $\epsilon_i$ through the
transformation of a group of independent random variables $\eta_l$
in terms of an identical Gaussian density distribution,
\begin{equation}
p_\eta (\eta)=(1/\sqrt{2\pi}\sigma)\exp[-\eta^2/2\sigma^2].
\end{equation}
Let
$q=e^{-1/l_{cor}}$,
the exponentially correlated variable
is generated implicitly by
$\epsilon_i={\displaystyle {\sqrt{1 - q^2}} \sum_{l=0}^{\infty} }
\eta_{i-l}q^l$, or equivalently in the following recursive form,
\begin{equation}\epsilon_L={\sqrt{1 - q^2}\eta_L} + q \epsilon_{L-1}.
\end{equation}
The three parameters $\lambda$, $\theta$, and $\epsilon$ at a step $L$
are what we need in order to calculate new parameters for the next step
$L+1$.

In the localized region, the equation we obtained for the
density distribution function $p(\theta,\epsilon)$ is
\begin{equation}
p(\theta,\epsilon)=\frac{1}{\sin^2\theta}
\int\mathrm{d}\eta \mathrm{d}\epsilon' \mathrm{d}\theta'
p_\eta(\eta) p(\theta',\epsilon')
\delta(\epsilon-{\sqrt{1 - q^2}\eta} - q \epsilon')
\delta({\frac 1 {\tan\theta}} + \tan\theta' -E +\epsilon).
\label{eq:theta_distri0}
\end{equation}
After we numerically solve this equation, the
Lyapunov exponent  $\gamma$ can be calculated through the following
formula,
\begin{equation}
\gamma={\frac 1 2 }
\int\mathrm{d}\eta \mathrm{d}\epsilon \mathrm{d}\theta
p_\eta(\eta) p(\theta,\epsilon)
\ln [1-(E-{\sqrt{1 - q^2}\eta} - q \epsilon)\sin 2\theta
+(E-{\sqrt{1 - q^2}\eta} - q \epsilon)^2\cos^2 \theta ].
\label{eq:length0}
\end{equation}
By defining the distribution function $p(\theta) $, which is
similar to the one in the uncorrelated diagonal disorder case,
\begin{equation}
p(\theta)=\int p(\theta,\epsilon)\mathrm{d}\epsilon,
\end{equation}
we obtain the same simple relationship between $p(\theta) $
and the Lyapunov exponent,
\begin{equation}
\gamma=-\int p(\theta)\ln |\tan \theta |\mathrm{d}\theta.
\label{eq:length}
\end{equation}

If we take the limit of $l_{cor} \to 0$ in the present correlated disorder situation,
the equations of $p(\theta)$ in the uncorrelated disorder case will
be recovered\cite{kkang2010,kkang2011}.  However, $p(\theta,\epsilon)$
is not exactly the product of $p(\theta)p_\eta(\epsilon)$ in this limit.

\begin{figure}
\includegraphics[width=12cm]{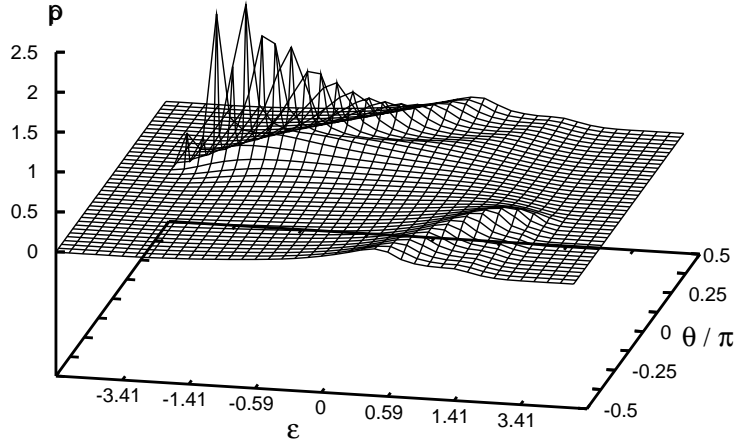}
\caption{Distribution $p(\theta,\epsilon)$ for $E=l_{cor}=\sigma=1$.
The forty lines in $\theta$ direction are evenly spaced in the region
$[-\pi/2,\pi/2]$. The forty lines in $\epsilon$ direction are scaled
to display a better global view.
} \label{fig:figure1}
\end{figure}

\begin{figure}
\includegraphics[width=12cm]{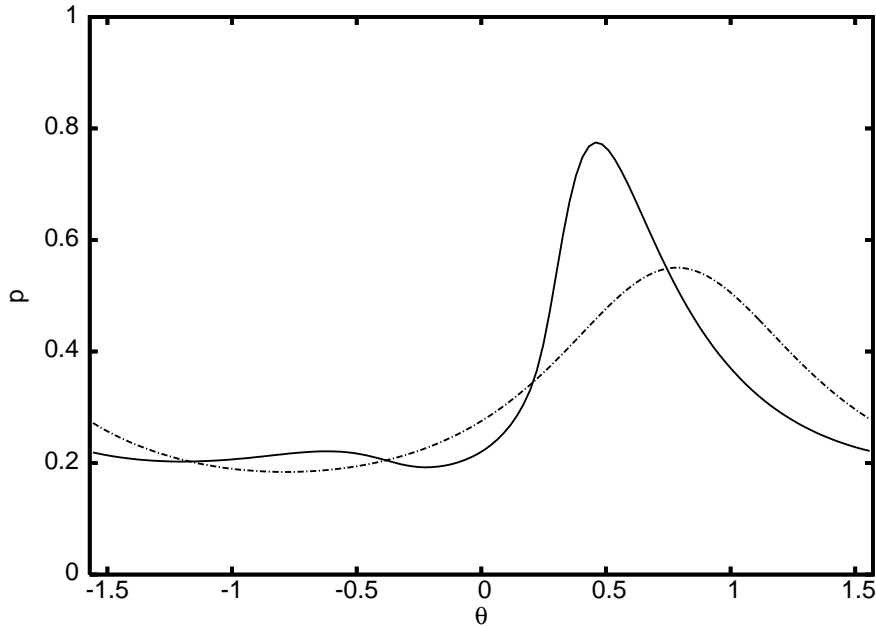}
\caption{Distributions $p(\theta)$ at $E=1$.  The full line is
for $l_{cor}=\sigma=1$; and dotted line for $l_{cor}=\sigma=0.01$.
Each point on the dotted line is differed
from the analytical curve  $\displaystyle {\frac { \sqrt{3} }{\pi(2-\sin 2 \theta)}}$
within no more than a relative error $10^{-4}$.
} \label{fig:figure2}
\end{figure}

We use the Gaussian distribution $p_\eta$ to solve
Eq.~(\ref{eq:theta_distri0}) and to calculate $\gamma$
numerically.  This method is very efficient to yield high precision results for
various disorder correlation length $l_{cor}$, disorder
strength $\sigma^2$, and energy $E$ in the thermodynamic limit.
Fig.~\ref{fig:figure1} and Fig.~\ref{fig:figure2} are shown
the calculated distribution functions $p(\theta,\epsilon)$
and $p(\theta)$, respectively.  In these calculations we have set a relative
precision $10^{-10}$ for $p(\theta,\epsilon)$.  Similar
distributions were calculated recently \cite{tkaya2009}
in the dichotomous correlated disorder case.

The structure of the joint density distribution of $\theta$
and $\epsilon$ is demonstrated by the $p(\theta,\epsilon)$
of $E=l_{cor}=\sigma=1$ in Fig.~\ref{fig:figure1}.  The
distribution is not so complicated to perceive, but it can not
be decomposed into a direct product of a density distribution
for $\theta$ and a density distribution for $\epsilon$.  Two
curves for $p(\theta)$ of $E=1$ are given in  Fig.~\ref{fig:figure2}.
One of the curve with $l_{cor}=\sigma=0.01$ has very small disorder
strength $\sigma$ and very small disorder correlation $l_{cor}$.
This curve can be approximated very well by the expression for
distribution $p(\theta)$ of uncorrelated disorder at a finite $E$ in
the weak disorder limit \cite{MKappus1981,FIzrailev1998,kkang2011},
\begin{equation}
p(\theta)=\frac{\sin\mu}{\pi(1-\cos\mu\sin 2\theta)}, \label{eq:uni},
\end{equation}
where $\cos \mu = E/2$.
Another curve with $l_{cor}=\sigma=1$ is not in the case for small disorder
strength or small disorder correlation, which is different from
the curve in small disorder strength or small disorder correlation.

The Lyapunov exponent  $\gamma(l_{cor},\sigma)$ is then calculated
by using the two curves shown in Fig.~\ref{fig:figure2} at $E=1$.  We obtain
$\gamma(1,1)=0.1252$ and $\gamma(0.01,0.01)=0.00001667$.
The direct calculated results from formula  Eq.~(\ref{eq:gmcor}) yield
$\gamma(1,1)=0.09587$
and $\gamma(0.01,0.01)=0.00001667$.  We see the numerically
calculated Lyapunov exponent for the finite energy and in the weak
disorder strength limit is well predicted by formula Eq.~(\ref{eq:gmcor}).

The situation for zero energy is different compared to that for the
finite energy.  There is no analytical result obtained so far for the zero
energy anomaly in the presence of correlated disorder in the weak disorder limit;
nor the formula predicting the Lyapunov exponent for a finite
correlation length. Our method is a good choice to perform calculation in these
situations.

\section{Anomaly at $E=0$}

We will investigate how the localization length changes
as the correlation of the disorder varies  at $E=0$ in the weak
disorder limit.  At finite energy, the disorder strength $\sigma$
and the correlation $\l_{cor}$ are decoupled
in function $\gamma(l_{cor},\sigma)$ in  Eq.~(\ref{eq:gmcor}).
Since  Eq.~(\ref{eq:gmcor}) was derived without any limitation on
the magnitude of the correlation length, with the help of the
Lyapunov exponent $\gamma(0,\sigma)$ for uncorrelated disorder,
the ratio
$\gamma(l_{cor},\sigma)/\gamma(0,\sigma)=\tanh{\frac{1}{2  l_{cor}}}$
might be exactly held for any $l_{cor}$.
At $E=0$ anomaly, even if we keep only $\sigma^2$ term in the weak
disorder limit, it is not known whether higher order terms from
correlation exists beyond the perturbation result.
To answer this question, we compare the numerically calculated result
with the perturbation one given by Eq.~(\ref{eq:gmcor}) at $E=0$:
\begin{equation}
\gamma_p = {\frac {\sigma^2 } {8 }}\tanh{\frac{1}{2  l_{cor}} }
\label{eq:gmcor0}.
\end{equation}

The small quantity related to correlation in $\gamma_p$ can be
considered in two limit cases. In the short correlation length limit
$l_{cor}\to 0$, i.e. $\tanh{\frac{1}{2  l_{cor}}} \to
 1$, the small quantity for expansion is $1-\tanh{\frac{1}{2 l_{cor}}}
\sim 2e^{-1/l_{cor}}$; whereas in the large correlation length limit
$l_{cor}\to \infty$, the small quantity for expansion is
$\tanh{\frac{1}{2  l_{cor}}}$ itself, $\tanh{\frac{1}{2  l_{cor}}}
\sim {\frac{1}{2  l_{cor}}}$. Therefore, we will calculate for a
group of different correlations with $\tanh{\frac{1}{2  l_{cor}}}$
close to zero as well as to one. In order to neglect the contribution from the
higher order terms of $\sigma$ in our calculation in the weak
disorder limit, we will calculate only for small $\sigma$.  It is sufficient to
keep three significant digits for the Lyapunov exponent $\gamma(
l_{cor},\sigma)$.

\begin{figure}
\includegraphics[width=12cm]{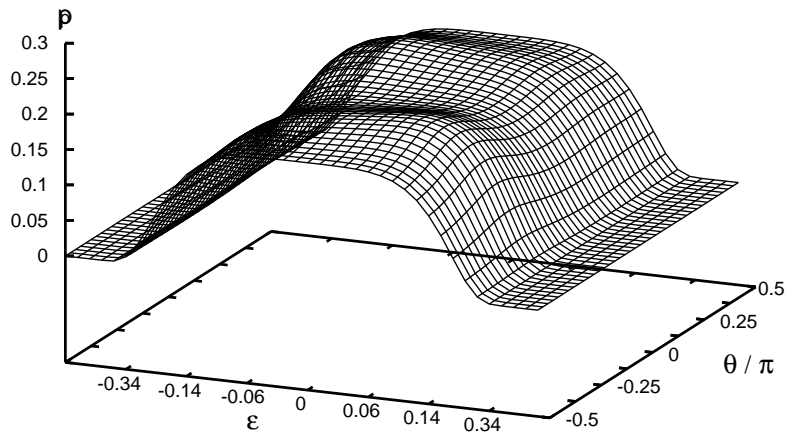}
\caption{Distribution $p(\theta,\epsilon)$ for $E=0$,
$\tanh{\frac{1}{2  l_{cor}}}=0.475$, and $\sigma=0.1$.
The forty lines in $\theta$ direction are evenly spaced in between
$[-\pi/2,\pi/2]$. The forty lines in $\epsilon$ direction are scaled
to give a better global view.
}
\label{fig:figure3}
\end{figure}

\begin{figure}
\includegraphics[width=12cm]{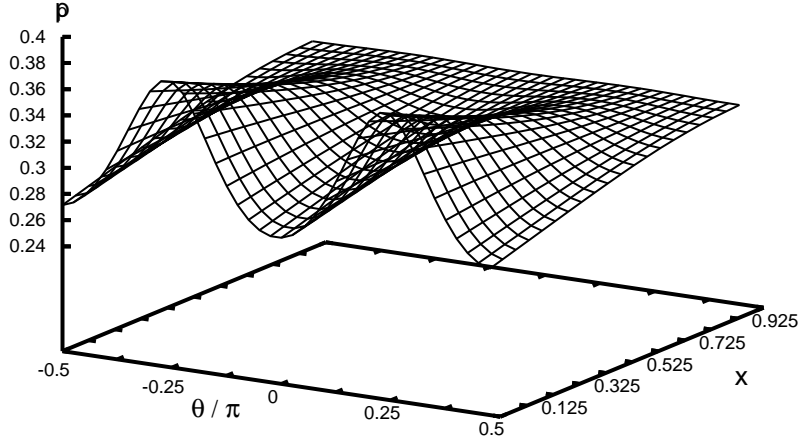}
\caption{Distributions $p(\theta)$ for $E=0$ and $\sigma=0.1$.
The forty lines in $\theta$ direction are evenly spaced in between
$[-\pi/2,\pi/2]$.  The twenty lines in $x$ direction are for
$x=1-\tanh{\frac{1}{2  l_{cor}}}=$ $0.025$, $0.075$, $0.125$, $\ldots$,
$0.925$, $0.975$, respectively.
} \label{fig:figure4}
\end{figure}

To demonstrate the anomalous behavior at $E=0$, we plot
in Fig.~\ref{fig:figure3} the distribution $p(\theta,\epsilon)$
for $E=0$, $\tanh{\frac{1}{2  l_{cor}}}=0.475$, and $\sigma=0.1$;
and in Fig.~\ref{fig:figure4} the distributions $p(\theta)$ for $E=0$
and  $\sigma=0.1$ with $x=1-\tanh{\frac{1}{2  l_{cor}}}=$ $0.025$,
$0.075$, $0.125$, $\ldots$, $0.925$, and $0.975$, respectively.
It shows clearly in Fig.~\ref{fig:figure3} that the joint distribution
for $\theta$ and $\epsilon$ has some inner structure.
We have observed the flattening of the distribution
$p(\theta,\epsilon)$ when increasing the correlation length $l_{cor}$
in the weak disorder limit.  The flattening will not be presented
in Fig.~\ref{fig:figure3}.

The flattening of $p(\theta)$ can be seen in Fig.~\ref{fig:figure4}.
In the figure, when correlation length is small, the
distribution  $p(\theta)$ turns out to be similar to the distribution for the
uncorrelated disorder\cite{CBarnes1990,FIzrailev1998,kkang2011}:
\begin{equation}
p(\theta)=\frac{1}{K(1/2)\sqrt{3+\cos4\theta}},
\end{equation}
where $K$ is the complete elliptic integral of the first kind.
As the correlation increases, we see that the distribution  $p(\theta)$
flattened towards $1/\pi$.  Let's take $x=1-\tanh{\frac{1}{2  l_{cor}}}$
as a new parameter of the correlation in disorder.
In the limit of $x \to 0$, which corresponds to the
uncorrelated limit $l_{cor} \to 0$, both the anomalous distribution
$1/\sqrt{3+\cos4\theta}$ and the anomalous Lyapunov exponent
$\gamma=\sigma^2/8.754$ of the uncorrelated disorder will be recovered.
In the limit of $x \to 1$, which is equivalent to the large correlation limit
$l_{cor}\to \infty$, $p(\theta)=1/\pi$ will correctly give
a zero Lyapunov exponent.

\section{High order terms of correlation at $E=0$}

Now we analyze the contribution from higher order terms of the
correlation in the weak disorder limit.  In Ref.~\cite{HSchomerus2003}
the authors gave analyses, which cover not only the localization length, but also
all the higher moments of the distribution of the Lyapunov exponent for
uncorrelated finite systems.  For correlated systems we expect
the deviation from Eq.~(\ref{eq:gmcor0}) comes from higher order
terms of correlation too.

\begin{figure}
\includegraphics[width=12cm]{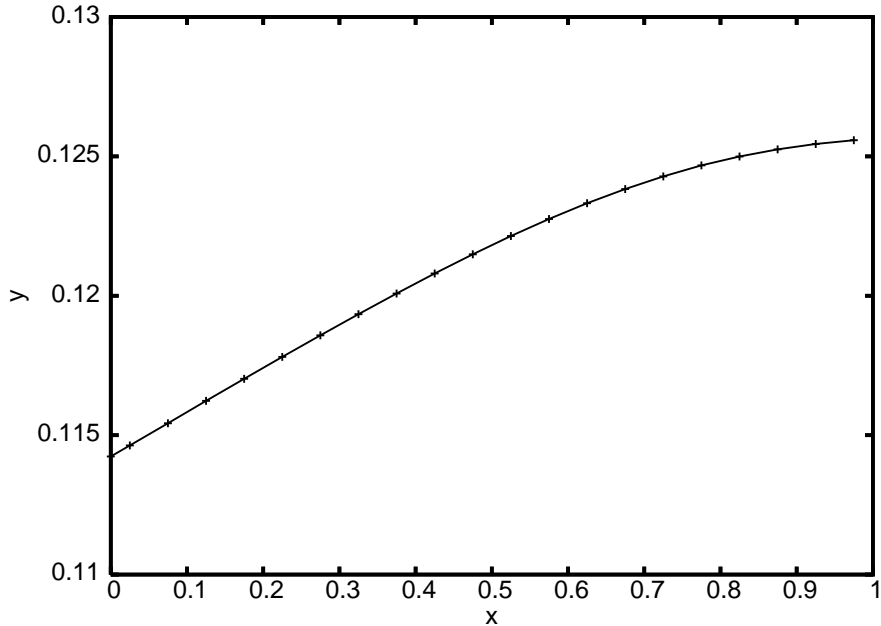}
\caption{The Lyapunov exponent $\gamma$ for $E=0$ and $\sigma=0.1$.
The variable $x$ used for different correlations is
$x=1-\tanh{\frac{1}{2  l_{cor}}}$.  The function is
$y=\gamma/(\sigma^2 \tanh{\frac{1}{2  l_{cor}}})$.
When $x$ is close to zero, $y$ is close to $1/8.754$;
and when $x$ is close to one, $y$ is close to $1/8$.
} \label{fig:figure5}
\end{figure}

In the perturbation result $\gamma_p$ in Eq.~(\ref{eq:gmcor0}),
by using variable $x=1-\tanh{\frac{1}{2  l_{cor}}}$ to denote the correlation,
we see that $\gamma_p$ included the first order correction of small
$x$ when $x\to 0$ and also the first order correction of small $1-x$
when $x\to 1$.  $\gamma_p$ has included only the
first order term.  From the discussion on $E=0$ anomaly in the previous section
we know that  $\gamma/(\sigma^2\tanh{\frac{1}{2  l_{cor}}}) = 1/8.754$
for $x\to 0$, while  $\gamma/(\sigma^2\tanh{\frac{1}{2  l_{cor}}}) = 1/8$
is predicted by perturbation result for $E\to0$.  The question on how
$\gamma$ really behaves at $E=0$ is still not answered:  whether
$\gamma/(\sigma^2\tanh{\frac{1}{2  l_{cor}}}) = 1/8.754$ always holds, or
there is a crossing to $\gamma/(\sigma^2\tanh{\frac{1}{2  l_{cor}}}) = 1/8$
as $ l_{cor}$ increases.  We plot  Fig.~\ref{fig:figure5} to answer
this question.

In  Fig.~\ref{fig:figure5}, the Lyapunov exponent $\gamma$ for $E=0$
and $\sigma=0.1$ are presented.  We plot for different correlations by
using the parameter $x=1-\tanh{\frac{1}{2  l_{cor}}}$, and we plot
$y=\gamma/(\sigma^2 \tanh{\frac{1}{2  l_{cor}}})$ as the function of $x$.
When $x$ is close to zero, $y$ is close to $1/8.754$; and when $x$
is close to one, $y$ is close to $1/8$.  We see a crossover between
the anomalous value $1/8.754$ and the perturbation result $1/8$.
In the weak disorder limit, besides the term $\tanh{\frac{1}{2  l_{cor}}}$,
there are higher order terms in $x$ or $1-x$ from the correlation.
The higher order terms connect smoothly the anomalous $1/8.754$
at zero correlation with the perturbation result $1/8$ for large
correlation length.

The physical picture is rich behind a finite magnitude of $\sigma$
and a large correlation length.  The $\sigma$ in Fig.~\ref{fig:figure5}
is not a small enough disorder strength.  The higher order terms
in $\sigma^2$ contributes when $x$ approaches one in
Fig.~\ref{fig:figure5}.  We have calculated
for much smaller $\sigma$ and confirmed that the contribution of higher
order terms in $\sigma^2$ goes to zero in the weak disorder limit.  Our
observation suggests further perturbation investigations.

\begin{figure}
\includegraphics[width=12cm]{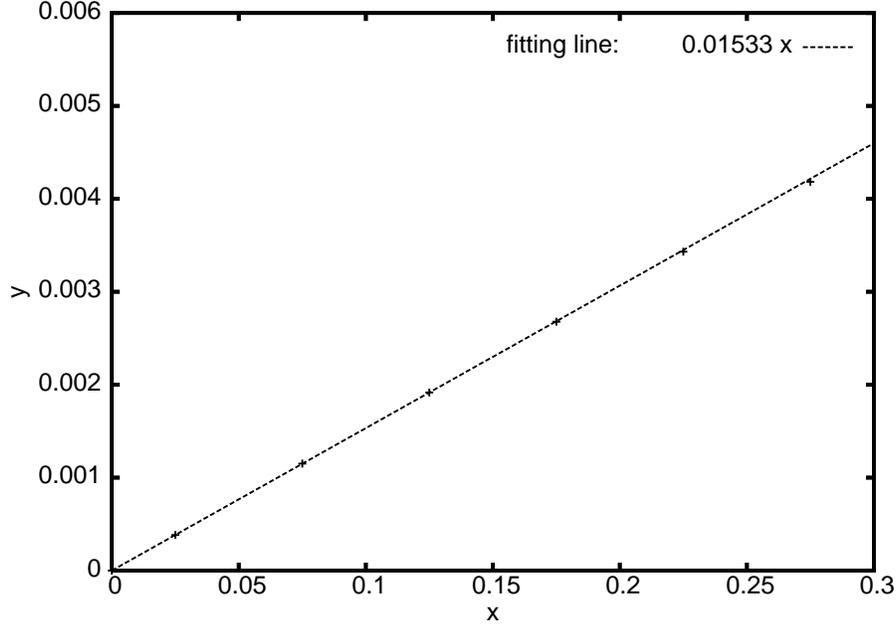}
\caption{Fitting of the Lyapunov exponent $\gamma$ for
$E=0$ and $\sigma=0.01$.  $y$ is certain the difference between the
Lyapunov exponent for finite correlation and zero correlation:
$y=\gamma(l_{cor},\sigma)/\sigma^2/\tanh{\frac{1}{2  l_{cor}}}-1/8.754$.
The variable $x$ used for different correlations is
$x=1-\tanh{\frac{1}{2  l_{cor}}}$.  The fitting line is $y = 0.01533 x$.
} \label{fig:figure6}
\end{figure}

To numerically provide the next leading term of the correlation
closed to the uncorrelated limit, we fit $\gamma$ for
$x$ close to zero in Fig.~\ref{fig:figure6}.  In Fig.~\ref{fig:figure6}
the Lyapunov exponent $\gamma$ for $E=0$ and $\sigma=0.01$ is plotted.
$y$ represents the difference between the Lyapunov exponent for a
finite correlation and for zero correlation:
$y=\gamma(l_{cor},\sigma)/(\sigma^2 \tanh{\frac{1}{2  l_{cor}}})-1/8.754$.
The variable $x$ used for different correlations is
$x=1-\tanh{\frac{1}{2  l_{cor}}}$.  We obtain a fitting line
$y = 0.01533 x$.  Therefore the perturbation expansion of $\gamma$
to the sub-leading order of the correlation in power of $x$ is obtained,
\begin{equation}
\gamma = ( 1 + 0.1342 x )
  {\frac {\sigma^2 } {8.754 }}
          \tanh{\frac{1}{2  l_{cor}} }
\label{eq:gm fit}.
\end{equation}
In Fig.~\ref{fig:figure5} it is clear that higher order terms contributes
when $x$ is even bigger.  In the weak disorder limit,
when $1-x$ close zero, the next order term in the correction
factor used to multiply to $\gamma_p$ in  Eq.~(\ref{eq:gmcor0}) is $(1-x)^2$.

\section{Conclusion}

In summary, we calculated the inverse localization length in
one-dimensional Anderson model with correlated diagonal disorder.
We obtained numerically the curve of the inverse localization length
for correlations at zero energy in the case of weak disorder. A
nonsingular curve was obtained for different correlation lengths in
the weak disorder limit at zero energy.

The variable used to plot the unifying curve is
$\tanh{\frac{1}{2  l_{cor}}}$, which has correspondence to
the Poisson process of the phase accumulation.   The inverse
localization length will be singular as the function of other
variables as $l_{cor}$, $1/l_{cor}$, or $e^{-1/l_{cor}}$.
We suggest further studies on the inverse localization length in
perturbation expansions or functional expansions
with the parameter $\tanh{\frac{1}{2  l_{cor}}}$.
We have obtained numerically in this work the next leading term
for comparison.

We also saw rich behavior for finite disorder strength and large
correlation length.  A unifying description of the band center
anomaly and the correlated disorder will be very interesting.

\begin{acknowledgments}
This work was supported by National Natural Science Foundation of
China No.~10374093,
and the Knowledge Innovation Project of Chinese Academy of Sciences.
\end{acknowledgments}

\bibliography{bib}

\end{document}